**Addressing the Challenges of Using Ferromagnetic Electrodes in the Molecular Spintronics Devices**

Pawan Tyagi[1, 2]*, Edward Friebe[1], Collin Baker[1]

[1]*University of the District of Columbia, Department of Mechanical Engineering, 4200 Connecticut Avenue NW Washington DC-20008, USA*

[2]*University of Kentucky, Department of Chemical and Materials Engineering 177 F. Paul Anderson Hall, Lexington KY-40506*

*Email of corresponding author\*: ptyagi@udc.edu*

**ABSTRACT:** Ferromagnetic (FM) electrodes chemically bonded with thiol functionalized molecules can produce novel molecular spintronics devices (MSDs). However, major challenges lie in developing FM electrodes based commercially viable device fabrication scheme that consider FM electrodes' susceptibility to oxidation, chemical etching, and stress induced deformations during fabrication and usage. This paper studies NiFe, an alloy used in the present day memory devices and high-temperature engineering applications, as a candidate FM electrode for the fabrication of MSDs. Our spectroscopic reflectance studies show that NiFe start oxidizing aggressively beyond ~90 $^0$C. The NiFe surfaces, aged for several months or heated for several minutes below ~90 $^0$C, exhibited remarkable electrochemical activity and were suitable for chemical bonding with the thiol functionalized molecular device elements. NiFe also demonstrated excellent etching resistance and minimized the mechanical stress induced deformities. This paper demonstrates the successful utilization of NiFe electrodes in the tunnel junction based molecular device fabrication approach. This paper is expected to fill the knowledge gap impeding the experimental development of FM based MSDs for realizing novel logic and memory devices and observing a numerous theoretically predicted phenomenon.

## 1. INTRODUCTION:

Realization of molecular spintronics devices (MSDs) for advanced computers will depend on successfully connecting molecules with the ferromagnetic (FM) electrodes [1-3]. However, device fabrication approaches may involve several steps that can oxidize,[4, 5] etch, and physically deform FM electrodes [6]. For example, oxidation may happen during exposure to air, high temperature photoresist baking during the photolithography step, annealing [4, 5], and exposure to oxygen (O) plasma [4], etc. Etching of FM electrodes may occur during interaction with process chemicals and molecular solutions. MSD can employ FM electrodes such as NiFe, cobalt (Co) and nickel (Ni). These FMs will have a tendency to oxidize [7, 8]. Already, Ni has been utilized in mechanical break-junctions [9] or electromigration-produced nanogap junctions [10]. To date, little attention has been given to NiFe as a FM electrode for the MSDs. NiFe alloys [11], which possesses better oxidation resistance as compared to Ni and iron (Fe) [12], have been widely utilized in commercial memory devices [5, 13, 14]. Due to thermodynamics NiFe surface contain Ni in the elemental form and Fe in the oxidized form [5]. Advantageously, NiFe also favors bonding with sulfur (S) over oxygen (O) [15]. Hence, thiol functionalized molecular device channels are expected to form stable chemical bonds. This paper discusses the utilization of a tunnel junction based molecular device (TJMD) approach to realizing the full potential of NiFe as the FM electrode for the MSDs. A TJMD approach utilizes a prefabricated tunnel junction with the exposed side edges (Fig. 1a) to host thiol functionalized molecular device channels (Fig. 1b) to yield a molecular device (Fig.1c). Distinctive advantages of TJMD approach over conventional molecular device fabrication approaches have been discussed elsewhere [1, 16]. However, to date no comprehensive study has been published on NiFe's merits and limitations pertinent to MSD fabrication. This paper investigate and discuss NiFe as a FM electrode and mainly utilized TJMD approach that is based on already commercialized magnetic tunnel junction (MTJ) technology [17].

## 2. EXPERIMENTAL METHODS:

This paper utilized Ni(80)Fe(20) in the form of unpatterned films and tunnel junctions (Fig. 1a). From this point onwards Ni(80)Fe(20) is referred as NiFe. For all the studies cleaned, oxidized silicon wafers were utilized. Thin film deposition was conducted with an AJA international multi-target sputtering machine. The deposition of NiFe and Co was accomplished with 150 W gun power at 2 mtorr argon gas pressure. The base pressure of the sputtering machine, before starting a deposition step, was $\sim 2 \times 10^{-7}$ torr. The deposition of nonmagnetic tantalum (Ta), generally used as a wetting layer, was deposited at 100 W gun power and 2 mtorr argon gas pressure. Details about NiFe based tunnel junction fabrication for molecular transport and magnetic studies have been published elsewhere [18, 19]. To study the changes in the NiFe surface, spectroscopic reflectance measurements were performed using the Semiconsoft M-Probe® system. We investigated the effect of temperature on the stability of ~10 nm thick NiFe deposited on ~2 nm Ta seed layer. We choose this NiFe thickness based on its typical thickness range in the recent TJMD [19] and memory devices[17]. A Ta (2 nm)/NiFe(10 nm) was heated from 22 $^0$C to 157 $^0$C temperature range; at every temperature setting a sample was heated for 30 minutes. For each temperature three different samples were studied; on each sample three scans were conducted. Before scanning the actual sample, a scan of reference oxidized silicon was used for calibration. Finally nine data sets were averaged to yield one representative data set for each temperature. We also conducted a gold (Au) metal deposition test on a NiFe surface to examine the changes in electrochemical activities with the exposure to ambient conditions. For the electrochemical studies, a Princeton Applied Research VersaSTAT 4 was utilized. The Au solution used for this study was TG-25 E and was purchased from TECHNIC INC, USA. We electrodeposited Au at ~ 10 mA/cm$^2$ current density. To verify the results of spectroscopic reflectance studies we also performed scanning tunneling microscopy (STM). For this study a NAIO STM with PtIr tips from Nanoscience Instrument® were utilized.

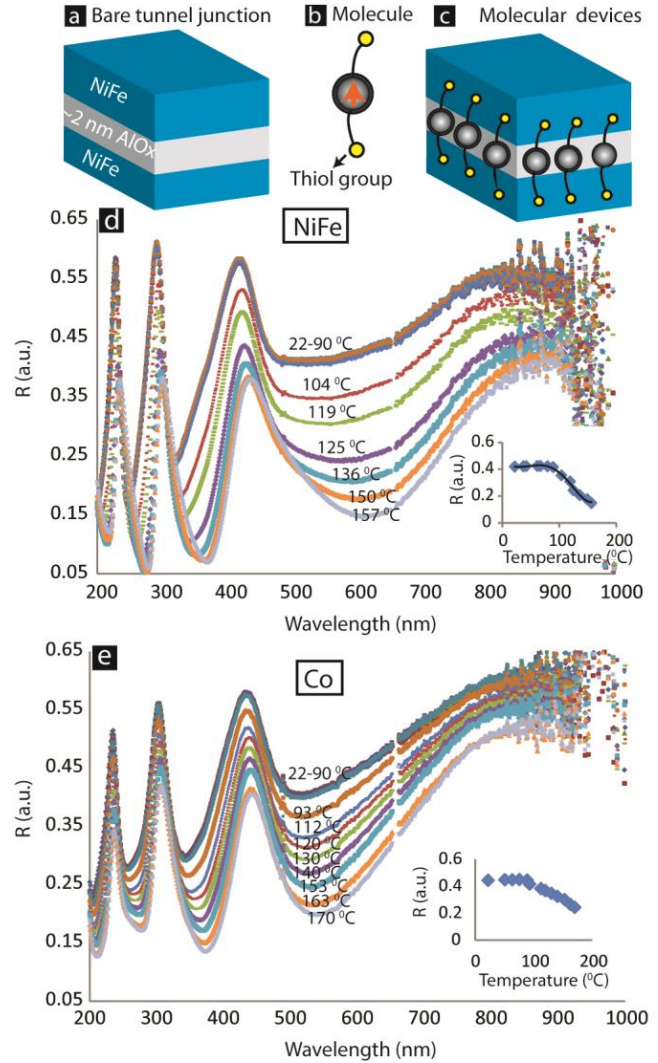

**Figure 1.** Scheme for including FM electrode in molecular device: (a) a prefabricated tunnel junction host (b) thiol functionalized molecules (c) by making thiol- FM metal chemical bonding. Spectroscopic reflectance measurement with temperature of (d) NiFe and (e) Co. Inset graph exhibits changes in the reflectance at 600 nm with temperature variation.

### 3. RESULT AND DISCUSSION:

To understand the oxidation NiFe tendency at different temperatures in the air, spectroscopic reflectance studies were conducted. NiFe did not exhibit any noticeable change in the reflectance between room temperature and ~90 $^0$C (Fig. 1d). Interestingly, after 90 $^0$C, the reflectance of NiFe started changing significantly (Fig. 1e). Change in reflectance around 600 nm wavelengths was very pronounced. The reflectance at 600 nm vs. temperature graph clearly shows that major changes in the NiFe occurred only around 90 $^0$C (inset graph of Fig. 1d). We also studied reflectance of the ~10 nm cobalt (Co) to investigate its oxidation characteristics with temperature. The Co layer showed that it is also stable up to ~90 $^0$C (Fig. 1e). The reflectance at 600 nm for Co also starts dropping around 90 $^0$C (inset graph of Fig. 1d). NiFe oxidation attributes appears to be similar to that of Co. However, elsewhere in this paper we discuss other criteria like chemical reactivity, and mechanical stability to mention the advantages of using NiFe over Co.

It is noteworthy that our reflectance vs. temperature data on NiFe was in excellent agreement with the Ferromagnetic Resonance (FMR) study on NiFe, subjected to oxidation conditions at different temperature [4]. According to Zhu et al.[4], the FMR line width (mT) vs. frequency graph showed no statistical difference between as grown and NiFe oxidized at 100 $^0$C; in this study temperature increased in 100 $^0$C step. Further heating changed the signal significantly. However, it must be noted the sample used in FMR study[4] possessed Ta(3 nm)/NiFe (20 nm)/AlOx (5 nm) configuration with AlOx on the top only allowing oxidation along the edges; on the other hand for our spectroscopic reflectance study Ta (2 nm)/NiFe(10 nm) with exposed top was utilized. Additionally, for the FMR study[4] oxidation was performed in a pure oxygen environment; however, in our case oxidation occurred in the air. These experimental differences may account for the difference in temperature for the onset start of the NiFe oxidation, 100 $^0$C for FMR and 90 $^0$C for the spectroscopic reflectance study (Fig. 1d). The reason behind sudden change in reflectance and FMR signal is presumably due to the changes in oxidation mechanism around 90 °C. At temperatures lower than 90 °C, the growth of an oxide film is presumably governed by the diffusion of metallic ions and electrons due to the influence of both concentration gradient and self-generated electric potential [12]. However, at higher temperatures oxidation kinetics is governed by the diffusion of O ions through the growing oxide layer [12].

In addition to the data discussed in Fig. 2, the published literature also provides useful insights about the NiFe surface interaction with O or air. Bruckner et al. [5] studied NiFe oxidation in air and corresponding changes in the surface chemistry. Experimental studies suggest that NiFe exhibited a preferred formation of FeOx over NiO; this result is consistent with observations made by Brundle et. al.[5] The preferred formation of FeOx is due to its lower free energy of formation than that of Ni. The free energy of formation of different oxides in kJ/mol are the following: FeO: -430, Fe$_2$O$_3$: -410, Fe$_3$O$_4$: -430, NiO: -350 [8]. Recently, Salou et al.[20] concluded that oxidation of NiFe is significantly less efficient than the oxidation of pure Ni and Fe surfaces [20]. The comparative XPS studies on polycrystalline Ni and NiFe showed that the effective thickness of NiO on NiFe was ~2 Å; while the effective thickness of NiO on pure Ni was ~12 Å [20]. This study substantiated that O dissolution into the NiFe was significantly less with respect to pure Ni. In addition, on the NiFe surface about 56% of the Fe was present as FeOx and hydroxide; however, on the pure Fe surface about 86% of Fe was present as FeOx and hydroxide [20]. The thickness of FeOx patches on the NiFe was determined to be 8 ± 2Å; which is significantly less than ~17 Å FeOx thickness observed on pure Fe [20]. Elsewhere in this paper we discuss that NiFe exposed to air is suitable for making molecular devices; however molecular device elements should preferably contain thiol functional groups (Fig. 1b) to bind with NiFe electrodes.

The effect of heating and exposing NiFe to ambient conditions was studied by other approaches to validate the interpretation from reflectance study (Fig. 1d). Visual inspection showed that the color of a Ta(2 nm)/NiFe (10 nm) film remained intact even after exposure to ambience for more than 10 months; the color of a Ta (2 nm)/NiFe(10 nm) is shown in (Fig. 2a). In another control study a Co(4 nm)/NiFe (6 nm) bilayer deposited on the oxidized silicon wafer did not show any noticeable change after aging for ~10 years in the open ambience; utility o Co/NiFe bilayer in MSD is discussed elsewhere in this paper. A fully oxidized NiFe or NiFe deposited on other thin films like Ta and Co exhibited dark blue color (Fig. 2b). The NiFe sputter deposited with O plasma also produced a blue color film. This observation showed that NiFe is stable and does not undergo progressive oxidation if stored near room temperature.

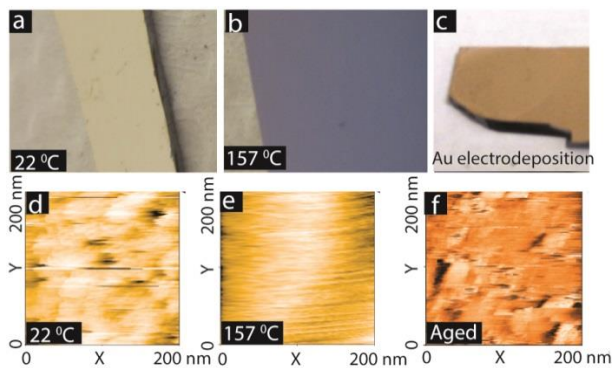

**Figure 2.** Ta(2 nm)/NiFe(10 nm) on the oxidized silicon in (a) as deposited state, (b) after heating at 157 $^0$C for 30 minutes, and (c) after electro deposition of Au. STM studies of Ta(2 nm)/ NiFe(10 nm) (d) at 22° C and (e) after heating at 157 ° C for 30 min. (f) STM of ~10 years old Co (4 nm)/NiFe (6 nm).

We investigated if surface oxides or adsorbed O on the NiFe surface has blocked the metallic sites and may prevent the bonding with the thiol functional group of a target molecule. Electrochemical reaction functionalization is a very effective way to form metal-thiol bonds [21]; however, a metal should be oxide free. Therefore we first attempted to ensure that the NiFe surface oxidized below 90 $^0$C is electrochemically active. We hypothesized that if NiFe is electrochemically active then it will respond to the electrodeposition of gold (Au) metal. We considered Au metal deposition because visual identification of Au is much easier than discerning monolayer(s) of electrochemically deposited thiol functionalized molecules. To validate this hypothesis we electrochemically deposited Au on Ta(2 nm)/NiFe(10 nm). NiFe surfaces of the samples were only rinsed with DI water prior to Au deposition. This step yielded visibly continuous Au film on NiFe surface and clearly showed that NiFe surface is not blocked by the oxide layer (Fig. 2c). The NiFe surface still allowed electrochemical Au deposition when heated at ~80 $^0$C for 5 minutes; however, heating NiFe at 157 $^0$C makes it unsuitable for electrochemical deposition. Ta/NiFe (top) bilayer color changed from silver white to deep blue due to oxidation. The ability to deposit Au is in agreement with the XPS studies emphasizing that NiFe exposed to air possesses conducting metallic Ni zones on the surface. We also hypothesized that sub-monolayer thick FeOx either may get etched off during electrochemical reactions or is conductive enough to enable Au deposition on it. This experiment suggests that NiFe has potential to electrochemically bind with thiol like terminal groups to make chemical bonds with the molecules of interest [22-25]. A prior study demonstrated that electrochemistry based self-assembly[21] was efficient and quick in establishing metal-molecule bonds [19].

To further test that NiFe surfaces do not possess a prohibitive oxide layer potentially hindering metal-thiol bonding, we conducted STM studies. STM study of few days old NiFe sample stored at room temperature showed good tunneling current and clear surface morphology. However, the NiFe sample heated at 157 $^0$C for 30 min showed the loss of tunneling current. The loss of tunneling current is due to the oxidation of Ta/NiFe layer; complete oxidation of the Ta/NiFe is also evident from the change in film color (Fig. 3a-b) and spectroscopic reflectance data (Fig. 2a). However, the tunneling current under the similar experimental conditions was still very high on the ~10 years old Co(4 nm)/NiFe (6 nm) surface. The order of tunneling current was in the nA range and was comparable to the freshly deposited NiFe. One direct implication of this experiment is that during molecular device fabrication the NiFe metal electrode must not be exposed to air at temperatures higher than 90 $^0$C; alternatively, one may ensure that inert gases are flushed on the sample during brief heating steps in photolithography-like processes.

For the development of novel MSDs it will be essential to change magnetic hardness of the two magnetic electrodes. Difference in magnetic hardness was crucial for observing molecule induced splitting of Kondo peaks in electromigration produced nanogap junction devices [3] and magneto resistance on magnetic tunnel junction based memory devices [17]. Moreover, tunnel junction based MSDs are better suited to explore the effect of varying magnetic hardness of FM electrode (Fig. 1a) as compared to the conventional MSD approaches [16]. In a magnetic tunnel junction one can easily manipulate NiFe magnetic attributes by depositing it with other magnetic or antiferromagnetic films [14, 17, 26]. However, in the event when NiFe is used along with other films there may be additional stability issues and concerns. For instance, a magnetic tunnel junction based molecular device utilizing a Co/NiFe bilayer as the bottom electrode and NiFe as the top electrode required the careful designing of the bottom electrode [6]. The ~10 nm thick bottom electrode with Co(2-8 nm)/NiFe (8-2 nm) were investigated for the application in molecular devices. According to SQUID magnetometer study the width of magnetic hysteresis loop of the Co (5 nm)/NiFe( 5 nm) was nearly four times more than that of NiFe (10 nm). We noted that in the event when Co thickness was 6-8 nm in the bilayer bottom electrode the tunnel junction testbeds were unstable, and also showed localized chemical etching during molecule attachment protocol (Fig. 3a). It is noteworthy that Co has a strong tendency to get etched when interacted with molecular solution during molecule attachment protocol (Fig. 3b). Here, 1-5 mM molecular solution was produced by dissolving organometallic molecular clusters (OMCs)[22, 27] in dichloromethane solvent. However, NiFe showed remarkable stability even after ~10 min of interaction with the molecular solution during molecule attachment protocol (Fig. 3c). Since in the Co/NiFe bilayer only Co is prone to etching we hypothesized that this metal is coming to the surface and showing localized pitting and etching. It is noteworthy that this phenomenon was mainly observed when Co thickness was >6 nm. We hypothesized that under the influence of compressive stresses a Co film moved vertically away from the substrate and poked through the NiFe top layer; this phenomenon only seems to occur with relatively thicker (>6 nm) Co because as film thickness increases its mechanical strength decreases.

To understand Co tendency to poke through the top layer we studied a more contrasting system consisting of a hard and air stable ~2 nm AlOx layer on the top of 10 nm Co. AlOx is a hard material and can protect materials underneath from ambience as well as display excellent strength under compressive load.

We were also interested in studying Co (10 nm)/AlOx (2 nm) bilayer for investigating the Co bottom electrode based MSDs. However, Co exhibited stress-induced mechanical deformations. We observed that on the Co/AlOx bilayer a number of nanoscale Co pillars popped out of the top AlOx surface; distinctive white color saturation regions on the 2D AFM images are due to Co pillars squeezed out through the hard AlOx layer (Fig. 3d). It was observed that depositing ~2 nm AlOx produced compressive stresses, which we hypothesized as the driving force behind the generation of nanoscale pillars (Fig. 3d). We believe that in the bilayer electrode the compressive stresses squeezed out Co through the weak spots in the NiFe top layer and interacted with acidic solution leading to the chemical etching (Fig. 3a).

We also measured the mechanical stresses in the sample and found that compressive stresses relaxed with time (Fig. 3e) leading to a concomitant increase in the population of as high as ~11 nm tall nanohillocks. In-depth discussion about the impact of stresses on the tunnel barrier stability has been discussed elsewhere [18]. We attempted several schemes for growing ~ 2 nm AlOx by depositing required amount of aluminum (Al) either in one step or in multiple steps, each step was followed by the plasma oxidation [18]. Each AlOx deposition scheme, involving deposition of 15 Å Al in single step, depositing 10 Å Al in the first step and 5 Å Al in the second step, and depositing Al in three steps of 5 Å each, produced compressive stresses which decreased with time (Fig. 3e). Analysis of AFM scan obtained on the Co/AlOx bilayer showed the presence of nanoscale hillocks of different height and diameter. This surface analysis was accomplished for all the schemes we considered for insulator growth. We finally selected that scheme for AlOx growth which produced low height of nanohillocks and low density [18].

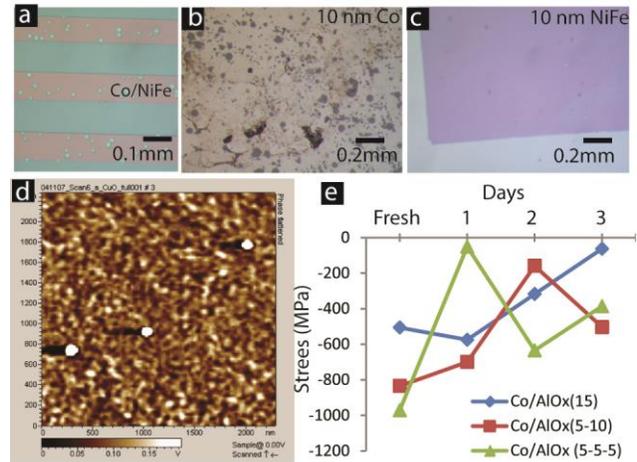

**Figure 3.** Exposure to dichloromethane-based molecular solution (a) damaged Co film but not (b) NiFe film. (c) Etching of stress gnererated Co pillars protruding out of NiFe in a NiFe/Co bilayer electrode. (d) the generation of Co pillars are seen as the spikes in the 3D graph. Control study showing (e) Time dependent relaxation of compressive stresses in AlOx covered Co film; AlOx was produced by different schemes.

The tendency of Co to form nanohillocks due to the mechanical stress relaxation signifies its unsuitability as a bottom electrode right under the ultrathin insulator and other NiFe like FM electrodes. To test Co metal electrode adaptation in prospective MSDs we produced exposed edge tunnel junction (Fig. 4a-e). In general, after the deposition of the first electrode (Fig. 4a) the photolithography step was performed to facilitate tunnel junction fabrication (Fig.4b-c) with the exposed side edges (Fig. 4d). During photolithography, NiFe of the bottom elecrode (Fig. 4a) remained under ~500 nm thick photoresist (PR) layer. This PR layer required 30-45 seconds long baking around 90 °C; keeping low baking temperature is justfied with the spectroscopic reflactance studies defining the safe temperature limits (Fig. 1d-e). After the photolithography step, 2 nm AlOx (Fig. 4b) and ~10 nm thick top metal electrode (Fig. 4c) were deposited [16]. After the liftoff of PR and undesired materials, we obtained a cross junction with the exposed side edges (Fig. 4d) where molecules could be placed between the two metal electrodes (Fig. 4e).

Our initial attempt to produce Co electrode based TJMDs showed unstable transport characteristics (Fig. 4). However, the tunnel junctions with ~ 10 nm Co bottom electrodes showed low leakage current in fresh conditions and was even smaller than the other two FM electrodes we studied (Fig. 4f and inset graph). But with time leakage current became quite unstable (Fig. 4h-j) and many tunnel junctions failed even without applying a breakdown voltage (Fig. 4h and j). It is apparent that nanohillocks generation (Fig. 3d), with time, created short-circuits between the two metal electrodes. We also noticed that, unlike NiFe, a Co electrode was prone to etching when exposed to some developer chemicals during photolithography. However, due to NiFe's chemical resistance, we experienced that Co(3-5 nm)/NiFe (7-5 nm) bilayer electrodes produced very stable and smooth FM electrodes and yielded low leakage current (Fig. 4f and g). The Rq roughness of Co/NiFe bilayer electrode was <0.4 nm rms. It is noteworthy that typical roughness of the commonly used ~10 nm Au electrode on the oxidized silicon wafer was ~1 nm. Depositing ~ 2 nm thick insulators for TJMDs is much easier on a smoother bottom electrode and this produces low leakage current. We studied the Co/NiFe/AlOx(2 nm)/NiFe tunnel junction transport at 50 mV under varying magnetic field, we observed magneto resistance; however, we were unable to see clear magnetoresistance step as seen from a well opimized magnetic tunnel junction [26].

We produced tunnel junction based molecular devices (TJMD) with NiFe electrodes. Exposed edges of a tunnel junction (Fig. 5a), produced after the photoresist (PR) liftoff (Fig,. 4d), were able to host OMC molecular channels (Fig. 5b) across the insulator to complete molecular devices fabrication (Fig. 5c). In this study we primarily used organometallic molecular clusters (OMCs) with an effective length of ~2.5 nm (Fig. 5b); details about protocol of OMC chemically bonding on a FM electrode [16, 19] and their synthesis [19, 22] are discussed elsewhere. The tunnel junction with NiFe (10 nm)/AlOx/NiFe (10 nm) exhibited a clear nonlinear tuneling transport chracterstics (Fig. 5d). Typically bottom and top electrode were 5 μm width, giving a tunnel junction area of 25 μm$^2$. This tunnel junction was subjected to cyclic voltage after submerging in ~2 mM OMC solution in $CH_2Cl_2$ solvent; current was recoded as voltage changed polaraity, typically between 0.1 V to -0.1 V, during this electrochemical step (Fig. 5e). Attachment of OMCs increased the current above the bare tunnel junction leakage current level (Fig. 5f).

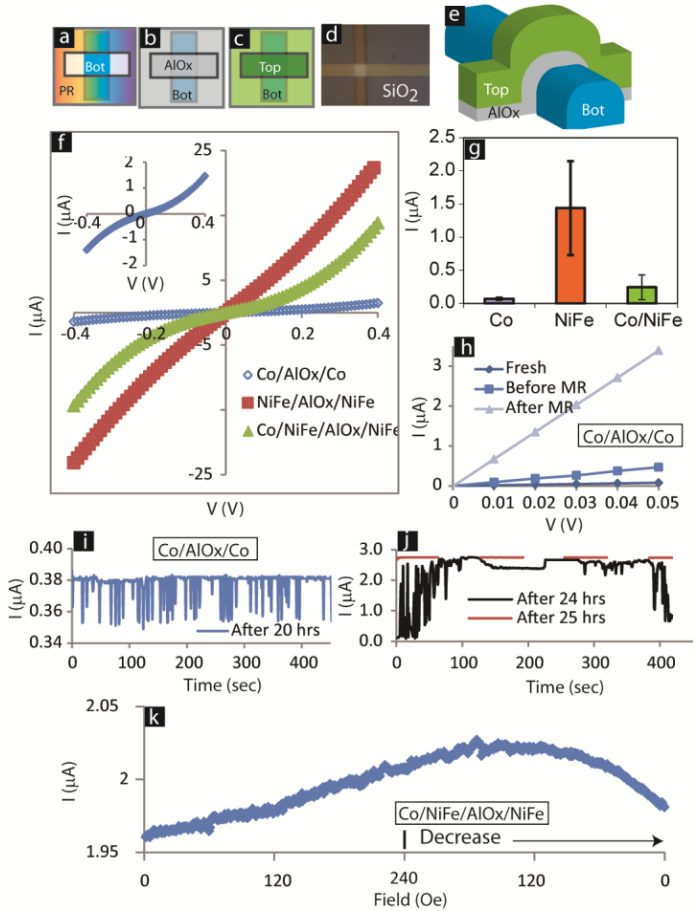

**Figure 4.** (a) photoresist cavity on a silicon wafer with the pre-deposited bottom (Bot) metal electrode. (b) Depositing 2 nm AlOx , (c) is folowed by the deposition of top metal electrode. (d) Micrograph and (e) 3D perspective view of a a tunnel junction after the liftoff step. (f) Current (I)-voltage (V) graph and (g) current at 50 mV for various tunnel junctions. (h) Co/AlOx/Co tunnel junction failure with time. Co/AlOx/Co showing (i) noisy current pattern at 50 mV which aggravates to (j) higher order noise in current and susequently failed. (k) A stable Co/NiFe/AlOx/NiFe tunnel junction showing change resistance at 50 mV with magnetic field.

All the successful TJMD were produced when FM films were sputter deposited in the photoresist cavity with slight undercut followed by the liftoff to yield bottom (Bot) electrode (Fig. 5g). The Co(3-5 nm)/NiFe(7-5 nm)/AlOx(2 nm)/NiFe(~10 nm) tunnel junctions produced with overall fabrication protocol discussed in Fig. 4a-e and Fig. 5g produced clear effect of OMCs. On the other hand tunnel junction with Co/NiFe bottom electrode produced by the chemical etching route (Fig. 5i) did not produced OMC effect (Fig. 5j). It was hypothesized that oxidation of FM electrodes ocuured during device fabrication that prevented OMC chemical bonding between two FM electrodes of the tunnel junction. During the hard baking of photoresist above 100 $^0$C for several minutes, typically required for high quality selective metal etching, the bottom FM electrode surface oxidized. It is noteworthy that right before baking photoresist was in lquid state and takes time to get hard. During this period when liquid resist is turning into solid the oxygen dose from air can oxidze top side of the bottom FM electrode (Fig. 5i). Oxidation of bottom FM electrode may worsen if the second photolithography step (Fig. 4a) also employ > 90 $^0$C baking temperature. However, ultra thin surface oxides may be taken care of by a mild chemical etching step of the completed tunnel junction (Fig. 4d); in electrodeposition of metals on Ni like FM metal a brief oxide removing etching is regularly utilized [28].

Prior research work on the NiFe alloys interacting with sulfur (S) containing gases in various engineering applications provides very useful insights; these insights can assist in understanding the viability and permanency of bonding between thiol functionalized molecules and NiFe. Here we assume that nature of the interaction of NiFe with the thiol (-SH) functional group is similar to that of S.

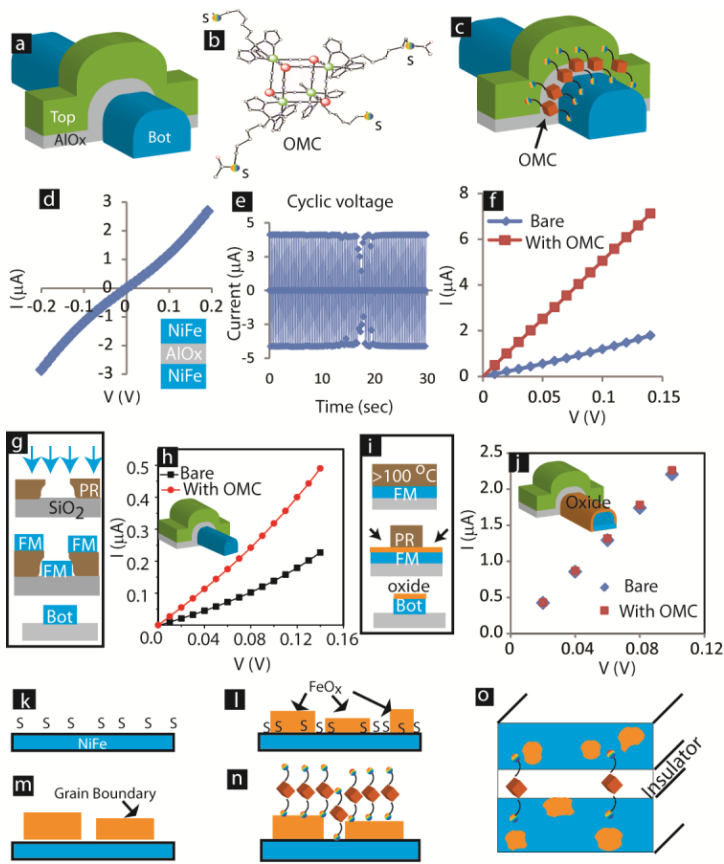

**Figure 5.** (a) A bare tunnel junction hosting (b) OMCs to become (c) molecular device. (d) I-V study of a bare NiFe/AlOx/NiFe tunnel junction. (e) Applying cyclic voltage on tunnel junction submersed in OMC solution to produce molecular bridges leading to (f) current increase. (g) Low temperature bottom (Bot) electrode deposition for tunnel junctions shown in panel (f) and (h). (h) OMC affecting bare Co/NiFe/AlOx/NiFe tunnel junction. (i) Bottom (Bot) produced by the selective etching of FM protected by hard baked PR for bare tunnel junction in panel (j) that did not show any sign of molecular device formation. (k) Pristine NiFe surface interacting with S. (l) Heating NiFe surface with S on it in the air lead to iron oxide (FeOx) formation while S remain chemisorbed. (m) Partly oxidized NiFe will have sub-nm thick FeOx pockets, surrounded by the elemental Ni pocket. (n) FeOx pockets preventing direct NiFe –molecule bonding. (o) Establishing molecular channels in the presence of surface oxides is less probable. possible.

Fotunately, S on NiFe is significantly more reactive and aggressive than O. In power generation plants S containing gases damages the NiFe [18, 29]. Sulfur was found to breakdown protective oxide layer and then chemisorbed on the NiFe surface [29]. According to Lad et al.[15, 29-31] due to favorable thermodynamics, S can persist on the NiFe surface for high temperatures in the O containing environments.

In a scenario where initially S chemisorbed on the pristine NiFe surface (Fig. 5k) and then oxidation commenced it was shown that subsequent oxide grew around S but, did not replace it with O (Fig. 5l). The binding energy of chemisorbed S on Ni and Fe is ~ 4.5 eV/atom [30]. However, the binding energy of O with Ni is ~3.8 eV/atom [32]. Due to this S can displace even adsorbed O on the NiFe surface and remain stable for >500 $^0$C [30]. However, if a NiFe surface is exposed to O and S together, it starts accommodating both. The equilibrium configuration of O and S affected surface area will be governed by both thermodynamic and kinetic factors [30]. During the fabrication of a NiFe based MSD some degree of surface oxidation is unavoidable. Partial oxidation is expected to lead to some sub-nm thick FeOx pockets that will be surrounded by elemental Ni on the surface (Fig. 5m) [5, 33]. FeOx can inhibit NiFe molecule bonding (Fig. 5n). The effect of surface oxide is limited to reducing the probability of bridging a large population of molecular channels between NiFe electrodes (Fig. 5o). This discussion strongly suggests minimizing ambient heating above 90 $^0$C to ensure minimal surface oxidation (Fig. 1)

### 4. COCLUSION:

MSD research can advance with the knowledge of suitable FM materials and methodology to incorporate them without adversely altering their surface properties. This paper discussed the NiFe surface oxidation properties with temperature in the air to understand the limits to be considered while designing fabrication schemes. We found that after the 90 $^0$C temperature, NiFe oxidation proceeded rapidly and hence suggests that device fabrication protocols should use this temperature as a threshold while performing a fabrication in the open ambience. However, we are not sure if this temperature limit is also good for the ambience with varying oxygen content and should be considered in the future studies. For the future attempts, we suggest that an inert gas or nitrogen environment be maintained to reduce the chances of FM electrode oxidation during baking. This paper also showed that NiFe is significantly resistant against chemical etching when exposed to common photoresist developer solutions utilized during photolithography, and dicholormethane like solvent used for molecular self-assembly. Molecules with thiol functional groups are an excellent choice for chemically bonding with the NiFe electrodes. NiFe bonds with S are highly stable in the ambient conditions and thermodynamically favored over oxidation. Considering the thermodynamic advantage, S can even replace a part of preexisting O on the NiFe surface and make chemical bonds. NiFe is also useful in mitigating the stress induced mechanical deformities on other common ferromagnetic materials like Co. Here we demonstrated TJMD device fabrication protocol, which was designed by carefully considering NiFe limits and merits.

**ACKNOWLEDGEMENT:**

A part of this study was supported by the National Science Foundation-Research Initiation Award (Contract # HRD-1238802), Department of Energy/ National Nuclear Security Agency subaward (Subaward No. 0007701-1000043016). We acknowledge the Air Force Office of Sponsored Research (Award #FA9550-13-1-0152) for facilitating study of ferromagnetic electrode stability by providing instrumentation support. PT thanks Dr. Bruce Hinds and the Department of Chemical and Materials engineering at University of Kentucky for facilitating experimental work on tunnel junction based molecular devices during his PhD. Molecules for molecular device fabrication were produced Dr. Stephen Holmes's group. Any opinions, findings, and conclusions or recommendations expressed in this material are those of the author(s) and do not necessarily reflect the views of any funding agency and corresponding author's past and present affiliations. The manuscript was written through contributions of all authors. Edward Friebe conducted spectroscopic reflectance measurements and electrochemical studies. Collin Baker performed STM studies.